\documentclass[prl,twocolumn,showpacs,preprintnumbers,amsmath,amssymb]{revtex4}

\usepackage{graphicx}
\usepackage{dcolumn}
\usepackage{bm}
\usepackage{isolatin1}

\def\eq#1{Eq.\ (\ref{#1})}
\def\eqs#1#2{Eqs.\ (\ref{#1}) and (\ref{#2})}

\def\tab#1{Table \ref{#1}}

\def\fig#1{Fig.~\ref{#1}}

\def\l{\left}
\def\r{\right}
\def\vec#1{\mathbf{#1}}

\def\msbar{{\overline{\mathrm{MS}}}}

\def\ri{{\mathrm{RI}}}
\def\rgi{{\mathrm{RGI}}}
\def\ndr{{\mathrm{NDR}}}
\def\pt{{\mathrm{PT}}}
\def\bare{{\mathrm{bare}}}
\def\refind{{\mathrm{ref}}}

\newcommand{\be}{\begin{equation}}
\newcommand{\ee}{\end{equation}}
\newcommand{\bea}{\begin{eqnarray}}
\newcommand{\eea}{\end{eqnarray}}

\newcommand{\nn}{\nonumber}
\newcommand{\ra}{\rightarrow}

\newcommand{\cR}{\mathcal{R}}

\newcommand{\cO}{\mathcal{O}}

\def\ord#1{O\l(#1\r)}

\newcommand{\la}{\langle}
\renewcommand{\ra}{\rangle}

\newcommand{\mev}{\mbox{\rm MeV}}
\newcommand{\gev}{\mbox{\rm GeV}}
\newcommand{\fm}{\mbox{\rm fm}}

\def\ods2{\mathcal{O}_{\Delta S=2}}
\def\zds2{Z_{\Delta S=2}}

\begin{document}

\preprint{CPT-2003/P.4545, CERN-TH/2003-142, BUHEP-03-14}

\title{$B_K$ from quenched QCD with exact chiral symmetry}

\author{Nicolas Garron$^1$, Leonardo Giusti$^{1,2}$, 
Christian Hoelbling$^1$, Laurent Lellouch$^1$ and Claudio Rebbi$^3$}
\affiliation{%
\vspace{0.2cm}
$^1$ Centre de Physique Théorique, Case 907, CNRS
Luminy, F-13288 Marseille Cedex 9, France\\
$^2$ Theory Division, CERN, CH-1211 Geneva 23, Switzerland\\
$^3$ Department of Physics, Boston University, 590
Commonwealth Avenue, Boston MA 02215, USA
}%

\date{June 30, 2003}

\begin{abstract}
We present a calculation of the standard model $\Delta S=2$ matrix
element relevant to indirect CP violation in $K\to\pi\pi$ decays which
uses Neuberger's chiral formulation of lattice fermions. The
computation is performed in the quenched approximation on a
$16^3\times 32$ lattice that has a lattice spacing $a\sim 0.1\,\fm$.  The
resulting bare matrix element is renormalized non-perturbatively. Our
main result is $B_K^\rgi=0.87(8)^{+2+14}_{-1-14}$, where the first
error is statistical, the second is systematic and the third is an
estimate of the uncertainty associated with the quenched approximation
and with the fact that our kaons are composed of degenerate $s$ and
$d$ quarks with masses $\sim m_s/2$.
\end{abstract}

\maketitle

{\em Introduction.---} Ginsparg-Wilson (GW) fermions
\cite{Ginsparg:1982bj,Kaplan:1992bt,Neuberger:1998bg,Neuberger:1998fp}, 
with their exact $SU(N_f)_L\times SU(N_f)_R$ chiral--flavor symmetry
at finite lattice spacing \cite{Luscher:1998pq}, provide unique
opportunities for exploring the physics of light quarks through
numerical simulations of lattice QCD (e.g. reviews in
\cite{Hernandez:2001yd,Giusti:2002rx}). In this regularization the
$\Delta S=1$ effective weak Hamiltonian renormalizes with the same
pattern as in the continuum. As a consequence, in the presence of an
active charm quark, the $\Delta I=1/2$ rule in $K\to\pi\pi$ decays can
be studied from the simpler $K\to\pi$ amplitudes without having to
perform difficult power subtractions \cite{Capitani:2000bm}.  In
addition, this exact chiral--flavor symmetry implies full
$\ord{a}$ improvement.  In this letter we present the results of a
quenched calculation of the matrix element of the $\Delta S=2$
effective weak Hamiltonian relevant to $K^0{-}\bar K^0$ mixing,
performed using Neuberger's implementation of GW
fermions~\cite{Neuberger:1998bg, Neuberger:1998fp}.  The bare matrix
element is renormalized non-perturbatively in the RI/MOM scheme à la
\cite{Martinelli:1995ty}.  This calculation is the first study of a
matrix element of a four-quark operator using a lattice formulation
which has exact chiral--flavor symmetry at finite lattice spacing. It
calls upon many of the ingredients required for studying the $\Delta
I=1/2$ rule or $\epsilon'$ (three-point functions, non-perturbative
renormalization of four-quark operators, etc.), but avoids the
numerically challenging computation of eye diagrams. Furthermore,
because this $\Delta S=2$ matrix element has been extensively studied
with other fermion formulations (e.g. reviews in
\cite{Battaglia:2003in,Lellouch:2002nj}), it provides a good test of
the reliability of our approach.

In addition to using Neuberger fermions, we make a number of other
improvements on the methods traditionally used to compute $\Delta S=2$
matrix elements.  In order to eliminate sizeable unphysical
finite-volume contributions from topology-induced fermion zero-modes,
we use the fact that the relevant $\Delta S=2$ operator is purely
left-handed and consider correlation functions constructed
from left-left quark propagators only. These propagators are free from
zero-mode contributions, as the latter are chiral. Thus, to create and
destroy kaons at rest, we use the time component of left-handed,
quark-bilinear currents. The use of left-handed currents was proposed
in \cite{Giusti:2002sm,Hernandez:2002ds} in the context of the
$\epsilon$-regime of Gasser and Leutwyler \cite{Gasser:1987ah}.  We
also use chiral sources in the RI/MOM non-perturbative renormalization
(NPR) of our operators to avoid zero-mode contributions, which would
otherwise appear. This has the added advantage of eliminating the leading
chiral-symmetry-breaking contributions to the NPR Green functions.

$K^0{-}\bar K^0$ mixing induces indirect CP violation in $K\to\pi\pi$
decays, quantified by the parameter $\epsilon$ for which the
theoretical expression~\cite{Buras:1998ra} and experimental
value~\cite{Hagiwara:2002fs} are:
\be
\epsilon\, e^{-i\frac{\pi}{4}}
\simeq C_\epsilon\, C_K(\mu)B_K(\mu)A^4\lambda^{10}\bar\eta\l[(1-\bar\rho) 
\eta_2S_0(x_t)\r.
\label{eq:eps}
\ee
$$
\l.+P(x_t,x_c,\ldots)\r]
= (2.282\pm 0.017)\times 10^{-3}
\ ,
$$
with $B_K$ defined through ($F_\pi=93\,\mev$)
\be
\la\bar K^0|\ods2(\mu)|K^0\ra=
\frac{16}{3}M_K^2\,F_K^2\times B_K(\mu)
\label{eq:ds2me}
\ee
and
\be
\ods2=[\bar s\gamma_\mu(1-\gamma_5) d]
[\bar s\gamma_\mu(1-\gamma_5) d]
\ .
\label{eq:rends2op}
\ee
Here, $C_\epsilon$ is obtained from well measured quantities, $A$,
$\lambda$, $\bar\rho$ and $\bar\eta$ are Wolfenstein parameters
\cite{Wolfenstein:1983yz} and $\eta_2$, $C_K$, $S_0$ and $P$ 
incorporate perturbative, short-distance physics ($P$ also contains
CKM factors). Combined with a determination of $B_K$, the measurement
of $\epsilon$ provides an important hyperbolic constraint on the
summit $(\bar\rho,\bar\eta)$ of the unitarity triangle, as indicated
by
\eq{eq:eps}. It is advantageous to compute $B_K$ instead
of the $\Delta S=2$ matrix element, 
because many lattice systematic errors cancel in the 
ratio that defines $B_K$.
A preliminary version of the present calculation was reported in
\cite{Garron:2002dt}. Preliminary results obtained by the MILC collaboration
with a different implementation of GW fermions were also published in
the same volume \cite{DeGrand:2002xe}.

\smallskip

{\em Computational details.---} The simulation is performed in
quenched QCD with $\beta=6.0$ and $V=16^3 \times 32$. We use a sample
of $80$ gauge configurations, generated with the standard Wilson
gluonic action, retrieved from the repository at the
``Gauge Connection'' (cf.~http://qcd.nersc.gov). Fermion propagators are
obtained from a local source, using Neuberger's action
\cite{Neuberger:1998bg,Neuberger:1998fp} with bare quark masses
$am=0.040,0.055,0.070,0.085,0.100$. The sign function of the Hermitian 
Wilson-Dirac operator, $X$, 
is obtained by making an optimal rational approximation
\cite{Neuberger:1998my,Edwards:1998yw,Edwards:1998wx}, after 
explicit evaluation of the
contributions from the lowest eigenvectors of $X^\dagger X$. The computation
of the propagators thus uses nested 
multi-conjugate gradient inversions (for more details
see~\cite{Giusti:2001yw,Giusti:2001pk}). The lattice spacing is 
determined from $r_0=0.5\,\fm$, where $r_0$ is the Sommer scale
\cite{Sommer:1994ce}. This gives $a^{-1}=2.12\,\gev$.

\smallskip

{\em Meson correlation functions and fits.---} From the quark
propagators we compute the two-point function
\be
C_{JJ}(x_0) =  \sum_{\vec x}\langle J_0(x) \bar J_0(0)\rangle 
\ ,\ee
and the three-point function
\be
C_{J\cO J}(x_0,y_0) =  
\sum_{\vec x,\vec y}\langle J_0(x) \ods2^\bare(0) J_0(y)\rangle 
\ ,\ee
where $J_\mu   =  \bar{d} \gamma_\mu (1-\gamma_5)\tilde s$, 
$\tilde q = (1-\frac{a}{2\rho} D)q$, 
$\rho=1.4$ (see \cite{Giusti:2001yw,Giusti:2001pk}), 
$\bar J_0$ is obtained from $J_0$ with $s\leftrightarrow d$
and $\ods2^\bare$ is obtained from \eq{eq:ds2me} with 
$d\to\tilde d$. Statistical errors are estimated with the jackknife method.

The ratio 
\be\label{eq:Ratio}
R(x_0,y_0) = \frac38\frac{C_{JOJ}(x_0,y_0)}{C_{JJ}(x_0) C_{JJ}(y_0)}
\ee
is fitted to the asymptotic form ($a\ll x_0\ll y_0\ll T$):
\be\label{eq:3ptfit}
R(x_0,y_0)\longrightarrow B_K^\bare
\ ,\ee
where  the fit parameter $B_K^\bare$  is the bare bag parameter.

The kaon mass $M_K$, on the other hand,
is determined from a fit of the time-symmetrized, two-point function 
to the standard asymptotic form ($a\ll x_0$)
\be
C_{JJ}(x_0) \longrightarrow \frac{Z^2}{M_K}e^{-M_KT/2}\cosh{\l[M_K(T/2-x_0)\r]}
\ ,\ee
with $Z=|\la K^0|J_0|0\ra|$. We find that the two-point function is
asymptotic from $ax_0=5$ on and that the statistically optimal fitting
range for $aM_K$ is $5\le ax_0\le 12,\,13,\,14,\,15$ and 16 when 
$am=0.040,\,0.055,\,0.070,\,0.085$ and $0.100$, respectively.

For the ratio of \eq{eq:Ratio}, we find that the asymptotic behavior
sets in for $ax_0\ge 5$ and $ay_0\le 27$.  Because our lattice is
rather short in the time direction, we have to worry about
time-reversed contributions.  Assuming that the time-reversed matrix
element is approximately the same as the forward one and that the
two-kaon energy is approximately $2M_K$ (i.e. that finite-volume
effects are not significant), then for $ax_0\le 6, ay_0\ge 26$ the
time-reversed contributions never exceed 1.5\% of the forward
signal. We also checked this explicitly by a fit which includes
time-reversed contributions and allows for finite-volume shifts on the
time-reversed matrix element and the two-kaon energy. We thus use the
range $5\le ax_0\le 6$ and $26\le ay_0\le 27$ to calculate our
observables. All fits are excellent and our results for $aM_K$ and
$B_K^\bare$ are summarized in
\tab{tab:fitres}. 

\begin{table}[htb]
\begin{center}
\begin{tabular}{ccc}
\hline\hline
$a m $ & $a M_K$ & $ B_K^\bare(a)$ \\
\hline
0.040  & 0.253(5) & 0.70(7) \\
0.055  & 0.288(4) & 0.71(6) \\
0.070  & 0.321(3) & 0.73(5) \\
0.085  & 0.352(3) & 0.74(4) \\
0.100  & 0.382(2) & 0.75(4) \\
\hline
\hline
\end{tabular}
\caption{
\label{tab:fitres}
Mesons masses and $B_K^\bare(a)$ {\em vs} quark mass.}
\end{center}
\end{table}

{\em Non-perturbative renormalization.---} 
We perform all renormalizations non-perturbatively in the RI/MOM
scheme à la \cite{Martinelli:1995ty}. Thus, we fix gluon configurations
to Landau gauge and numerically compute appropriate, amputated forward
quark Green functions with legs of momentum $p\equiv
\sqrt{p^2}$. We define the ratio
\be
\label{eq:zbkdef}
\cR^\ri(m,p^2,g_0)=\frac{\Gamma_J(m,p^2)^2} {\Gamma_{\ods2}(m,p^2)} 
\ ,
\ee 
where $\Gamma_\cO$ is the value of the non-perturbative, amputated
Green function of operator $\cO$ computed with left-handed quark
sources and projected onto the spin-color structure of $\cO$. We
compute this ratio for all five quark masses and perform a linear
chiral extrapolation in $m^2$ as prescribed by a next-to-leading order
expansion of the Green functions in quark mass. We find that the
mass dependence is very mild and is well described by this linear
form. We must then isolate the ``perturbative part'' of this ratio to get
the renormalization constant appropriate for renormalizing
$B_K^\bare(a)$. A straightforward operator product expansion (OPE) in
$1/p^2$ yields the following $p^2$ behavior for $\cR^\ri$:
\bea
\cR^\ri(0,p^2,g_0)&=&\cdots+\frac{A}{p^2}+Z_{B_K}^\rgi(g_0)
\times U^\ri(p^2)\nn\\
\label{eq:zbkope}
&&+B\times(ap)^2+\cdots
\ ,\eea
where $Z_{B_K}^\rgi(g_0)$ is the renormalization constant
that takes $B_K^\bare(a)$ to the renormalization-group invariant
parameter $B_K^\rgi$, and $U^\ri(p^2)$ describes the
running in $p$ of the corresponding RI-scheme renormalization constant
$Z_{B_K}^\ri(p^2,g_0)$. The ellipses in \eq{eq:zbkope}
correspond to higher-order terms in the OPE  
and higher-order discretization errors.

To describe the running of $Z_{B_K}^\ri(p^2,g_0)$, we use the
2-loop expression obtained by combining the $\msbar{-}\mathrm{NDR}$
result of \cite{Buras:1990fn} and the $\msbar{-}\mathrm{NDR}\to\ri$
matching result of \cite{Ciuchini:1995cd}, with $N_f=0$.  Thus,
\be
\label{eq:2loop running}
U^\ri(p^2)=\alpha_s(p^2)^{\frac{\gamma_0}{2\beta_0}}
\l[1+\frac{\beta_1}{\beta_0}\frac{\alpha_s(p^2)}{4\pi}\r]^{\frac{\gamma_1^\ri}{2\beta_1}
-\frac{\gamma_0}{2\beta_0}
}
\ ,\ee
with $\beta_0=11$, $\beta_1=102$, $\gamma_0=4$ and $\gamma_1^\ri=287/3-176
\ln 2$. The strong coupling constant that we use is taken from \cite{Capitani:1998mq}
and corresponds to $\alpha_s(\mu_\refind^2)=0.0926(12)$ in the $\msbar$
scheme with $\mu_\refind=94.1(3.6)/r_0$. 
To obtain the coupling at other scales, we integrate the 2-loop running equation
exactly and solve it numerically.

\begin{figure}
\vspace{0.5cm}
\includegraphics*[width=8cm]{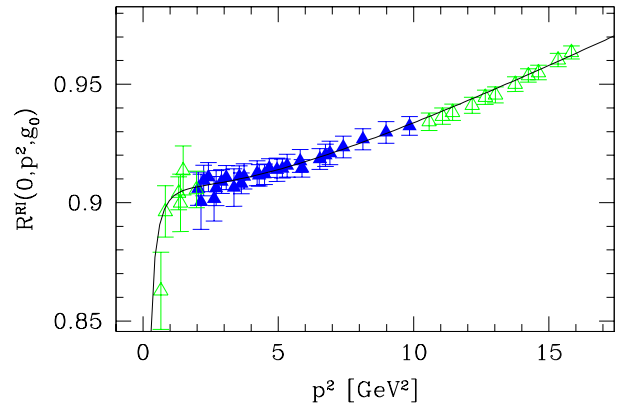}%
\caption{\label{fig:rri}$\cR^\ri(0,p^2,g_0)$ {\em vs} $p^2$. The solid
triangles correspond to the points used in the fit to \eq{eq:zbkope} while the open ones
are not included. The curve is the result of the fit.}
\end{figure}

Our results for $\cR^\ri(0,p^2,g_0)$ are plotted in
\fig{fig:rri} as a function of
$p^2\equiv\frac{(a/r_0)^2}{a^2}\sum_{\mu=0}^{3}(\sin ap_\mu)^2$. Using
the lattice definition of momentum significantly reduces
discretization errors.  Also shown is the fit of these results
to the functional form given by \eqs{eq:zbkope}{eq:2loop running} with
$Z_{B_K}^\rgi(g_0)$, $A$ and $B$ as parameters, in the range of
$2\,\gev^2\le p^2\le 10\,\gev^2$. We find that the OPE and
discretization error terms kept in \eq{eq:zbkope} are sufficient to
describe the data in this range. The fit actually describes the data
in a much larger range, indicating that the retained terms 
dominate. The results of the fits are summarized 
in \tab{tab:zbkfitres}, where we also display a fit to the extended momentum
range $0.65\,\gev^2\le p^2\le 15.9\,\gev^2$ with additional
$1/p^4$ and $(ap)^4$ terms, and a fit using the continuum $p^2$ 
in the range $2\,\gev^2\le p^2\le 10\,\gev^2$. All fits are excellent
and produce compatible results.
The renormalization constants at $4\,\gev^2$ in the RI and
$\msbar{-}\mathrm{NDR}$ schemes are obtained by multiplying
$Z_{B_K}^\rgi(g_0)$ by the appropriate two-loop running expressions
$U^\ri(4\,\gev^2)$ and $U^\ndr(4\,\gev^2)$, respectively, with $N_f$ and
$\alpha_s$ chosen as above. $U^\ndr(p^2)$ is given by \eq{eq:2loop
running} with $\gamma_1^\ri\to \gamma_1^\ndr=-7$ for $N_f=0$.

\begin{table*}[htb]
\begin{center}
\begin{tabular}{lccccc}
\hline\hline
$p^2$-range $[\gev^2]$ ($p^2$ def.) \ \ \ 
& $Z_{B_K}^\rgi(g_0)$ & $1/p^2\;[\gev^2]$ & $(ap)^2$ & $1/p^4\;[\gev^4]$ & $(ap)^4$\\
\hline
$[2,10]$ (lattice) & 1.261(9) & $-0.042(16)$ & 0.028(2) & 0 & 0\\
$[0.65,15.9]$ (lattice) & 1.264(13) & $-0.035(23)$ & 0.024(5) & $-0.013(13)$ & 0.0010(9)\\
$[2,10]$ (continuum) & 1.269(8) & $-0.051(16)$ & 0.022(2) & 0 & 0\\
\hline
\hline
\end{tabular}
\caption{
\label{tab:zbkfitres}
Rernomalization constants and OPE coefficients obtained from fits of
$\cR^\ri(0,p^2,g_0)$ to the OPE form of \eq{eq:zbkope} for different
momentum ranges and for the lattice and continuum definitions of
$p^2$.}
\end{center}
\end{table*}

\smallskip

{\em Physical results.---} The central values for $B_K^\bare(a)$,
$Z_{B_K}^\rgi(g_0)$, $Z_{B_K}^\ndr(4\,\gev^2,g_0)$ and
$Z_{B_K}^\ri(4\,\gev^2,g_0)$ are obtained as described above. While
statistical and systematic errors in the $B$-parameters and
renormalization constants will be correlated in our final results, we
also wish to give results for the renormalization constants
themselves. The systematic errors on these constants take into account
the following sources of uncertainty: the errors on
$\alpha_s(\mu_\refind^2)$ and $\mu_\refind$, given above; the
variation due to a $\pm 10\%$ uncertainty on the lattice spacing,
which is typical in quenched calculations \cite{Aoki:2002fd}; the
difference between the renormalization constants obtained using the
lattice and continuum definitions of quark momentum. The latter yields
an estimate of discretization errors which turns out to be the
dominant uncertainty. We take it to be a symmetric error. Adding all
of these variations in quadrature, we obtain:
\bea
Z_{B_K}^\rgi(g_0)&=&1.261(9)^{+11}_{-10}\nn\\
\label{eq:zbkres}
Z_{B_K}^\ndr(4\,\gev^2,g_0)&=&0.908(6)^{+7}_{-6}\\
Z_{B_K}^\ri(4\,\gev^2,g_0)&=&0.897(6)^{+7}_{-6}\ ,\nn
\eea
for $\rho=1.4$. For comparison, we compute the renormalization
constant $Z_{B_K}^{\ri,\pt}$ at one-loop in perturbation theory,
matching the results of \cite{Capitani:2000da} with the standard
RI/MOM scheme used here. We find
$Z_{B_K}^{\ri,\pt}(4\,\gev^2,g_0)=0.95(5)$ with the value of
$\alpha_s$ given above. The central value is obtained by expanding the
ratio of $\zds2^{\ri,\pt}(4\,\gev^2,g_0)$ to $Z_{A}^{\pt}(g_0)^2$ in
$\alpha_s$. The uncertainty is determined from the errors on
$\alpha_s(\mu_\refind^2)$ and $\mu_\refind$ and from an estimate of
$\ord{\alpha_s^2}$ corrections obtained by considering the unexpanded
ratio.  While the one-loop correction to
$\zds2^{\ri,\pt}(4\,\gev^2,g_0)$ is very large (of order 70\%) and the
perturbative result for $Z_{A}^{\pt}(g_0)^2$ is approximately 15\%
below the non-perturbative value found in
\cite{Giusti:2001yw}, the agreement of
$Z_{B_K}^{\ri,\pt}(4\,\gev^2,g_0)$ with the non-perturbative value
given in \eq{eq:zbkres} is surprisingly good. This is due to the large
cancellations in $\zds2^{\ri,\pt}(4\,\gev^2,g_0)/Z_{A}^{\pt}(g_0)^2$.

\smallskip

\begin{figure}
\vspace{0.5cm}
\includegraphics*[width=8cm]{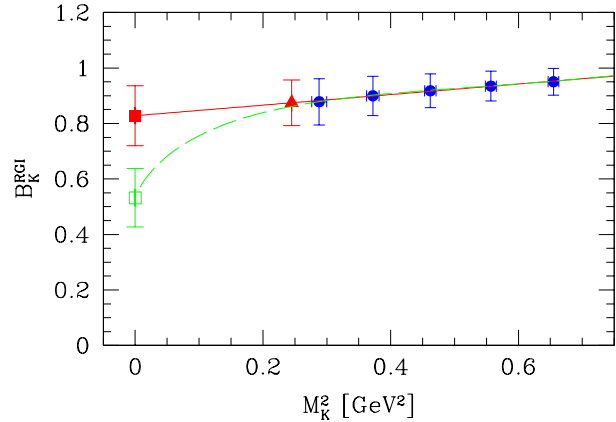}%
\caption{\label{fig:bkchi}The renormalization group invariant 
parameter $B_K^\rgi$
as a function of kaon mass squared. The solid circles are 
the result of our simulation which
are fitted to a line (solid curve) and the Q$\chi$PT 
expression (dashed curve). The
solid triangle is the value of $B_K^\rgi$ at the physical point. 
The squares correspond
to the chirally extrapolated values $B^\rgi$ as obtained 
from the linear (solid square) and
Q$\chi$PT (open square) fits.}
\end{figure}

We now turn to $B_K$. Our lightest pseudoscalar meson is very close to
having the mass of the kaon. Since our results for $B_K$ are linear in
$M_K^2$, we extrapolate them linearly to the physical point, as shown
in \fig{fig:bkchi}. To extrapolate to lighter quarks, chiral
logarithms should be taken into account.  Thus, we also perform a fit
to the functional dependence predicted by one-loop quenched chiral
perturbation theory (Q$\chi$PT)~\cite{Sharpe:1992ft}, supplemented by
an $\l(M/4\pi F\r)^4$ term to parametrize higher-order
contributions, which are expected for our pseudoscalar meson
masses. Here, $F$ is the leading-order leptonic decay constant, which we take to
be $F_\pi$, and $M$ is the leading-order pseudoscalar
meson mass, which we set equal to meson masses obtained in our
simulation. The fit parameters are the RGI value of the chiral-limit
$B$-parameter, $B^\rgi$, the scale of the one-loop logarithm and the
coefficient of the $\l(M/4\pi F\r)^4$ term. The fit,
also shown in~\fig{fig:bkchi}, is only meant to indicate how a chiral
extrapolation could lead to much smaller values of the $B$-parameter
in the chiral limit, such as the one found in the large-$N_c$
calculation of
\cite{Peris:2000sw}. We would need results at lower masses and
with better statistics to confirm the presence of the logarithm. For
the time being we find, in the quenched approximation,
$B^\rgi=0.53(11)^{+4+30}_{-3-0}$, where the central value is obtained
from the $\chi$PT fit. The first error here is statistical, the second
results from implementing the systematic variations that were
discussed above and the third is the difference between the central
values of linear and $\chi$PT fits.

As evident from \fig{fig:bkchi}, the value of $B_K^\rgi$ at the
physical point is insensitive to the choice of functional form
in the fit. We thus use the simpler linear fit to determine 
$B_K^\rgi$ at that point. We allow for a 20\% uncertainty in
the determination of the strange quark mass, which is typical of the
variations observed in quenched calculations of this quantity
\cite{Aoki:2002fd}. In addition, we account in a
correlated fashion for the statistical errors and the systematic
variations that were discussed in the paragraph on the
renormalization constants. We do not include a discretization error on
the bare value of $B_K$, because we have no way of estimating
it from our simulation performed at a single value of the lattice
spacing. However, $B$-parameters are ratios of very similar matrix
elements, in which some discretization errors should cancel. Moreover,
the limited experience that we have with Neuberger fermions
suggests that discretization errors on quantities such as $F_K$ are
small at $\beta=6.0$~\cite{Hernandez:2001hq}.

Our results for $B_K$ in various schemes are summarized in
\tab{tab:bkresults}. They are in excellent agreement
with world averages for this quantity (e.g. 
\cite{Battaglia:2003in,Lellouch:2002nj}), which are based on 
quenched, staggered results
\cite{Kilcup:1990fq,Kilcup:1998ye,Aoki:1998nr}. However, our
precision is not sufficient to exclude the rather low
values found with domain-wall fermions
\cite{AliKhan:2001wr,Blum:2001xb}.

\begin{table}[htb]
\begin{center}
\begin{tabular}{lccc}
\hline\hline
& RGI & NDR $(4\,\gev^2)$ & RI $(4\,\gev^2)$ \\
\hline
$B_K$ & $0.87(8)^{+2}_{-1}$ & $0.63(6)^{+1}_{-1}$ & $0.62(6)^{+1}_{-1}$\\
\hline
\hline
\end{tabular}
\caption{
\label{tab:bkresults}
Final results for $B_K$ at the physical point and in the chiral limit
in different schemes. The first error is statistical and the second is
systematic (see text for a discussion of additional $m_s=m_d$ and
quenching systematics).}
\end{center}
\end{table}

In addition to the systematic errors accounted for in the results of
\tab{tab:bkresults}, one must also consider the error associated with
the fact that our kaon is composed of degenerate quarks with masses
$\sim m_s/2$, instead of an $s$ and a $d$ quark. This error is thought
to be roughly 5\% on the basis of $\chi$PT estimates reviewed in
\cite{Sharpe:1998hh}. In this same review, Sharpe suggests adding a 15\%
quenching uncertainty. We have accounted for part of this error in
considering the variations due to the uncertainties in the
determination of the lattice spacing and of the strange quark mass. We
find these variations to be very much smaller than 15\%, which might
suggest that the quenching error estimate given in
\cite{Sharpe:1998hh} is rather conservative.

\section*{Acknowledgments}
We thank P. Hernández, M. Lüscher, M. Testa, P. Weisz and H. Wittig
for interesting discussions. We wish to thank Boston University's
Center for Computational Science and its Office of Information
Technology for generous allocations of supercomputer time, and the
Scientific Computing and Visualization group for invaluable technical
assistance. This work is supported in part under DOE grant
DE-FG02-91ER40676 and by the European Community's Human Potential
Programme under contracts HPRN-CT-2000-00145 (Hadrons/Lattice QCD) and
HPRN-CT2002-00311 (Euridice).  C.H. is supported by EU grant
HPMF-CT-2001-01468.

\vfill

\bibstyle{unsrt}
\bibliography{bk}


\end{document}